\documentclass[10pt,a4paper]{article}
\usepackage{amsmath}
\usepackage{amssymb}
\usepackage{amsfonts}
\usepackage{amstext}
\usepackage{amsbsy}
\usepackage[mathscr]{eucal}
\usepackage{graphicx}
\usepackage{color}
\usepackage[all]{xy}
\newcommand{\beq}{\begin{equation}}
\newcommand{\eeq}{\end{equation}}
\newcommand{\beqa}{\begin{eqnarray}}
\newcommand{\eeqa}{\end{eqnarray}}
\newcommand{\ket}[1]{| #1 \rangle}

\newtheorem{thm}{Theorem}[subsection]
 \newtheorem{cor}[thm]{Corollary}

 \numberwithin{equation}{subsection}

\makeindex
\title{\Large\textbf{Selective phase rotation quantum gate entangler}}

\author{\textit{ Hoshang Heydari}\\
        \small\textit{Department of Physics, Stockholm university 10691 Stockholm Sweden}\\
\\\small\textit{Email: hoshang@physto.se}}

\date{}
\begin{document}

\maketitle \thispagestyle{empty}

\begin{abstract}
We construct a quantum gate entangler based on selective phase
rotation transform. In particular, we established a relation between
quantum integral transform and quantum gates entangler in terms of
universal controlled gates for multi-qubit states. Our construction
is also related to the geometrical structure of multipartite quantum
systems.
\end{abstract}

\section{Introduction}
 In quantum computing we are interested in manipulated multipartite
 quantum states by quantum gates. One very important types of such
 gates are called quantum gate entangler which creates entangle
 states from product states. Recently, topological quantum gate
 entangler for two-qubit state has been constructed in \cite{kauf}.
  These topological operators are called braiding operators that can entangle
   quantum states and are also unitary solution of Yang-Baxter equation.
   Thus, topological unitary transformations are very suitable for
   application in the field of quantum computing.  Recently, we have also
   have constructed such quantum gate entangler for multi-qubit states and
   general multipartite states \cite{Hosh1,Hosh2}. In this paper we will
   construct quantum gate entangler based on unitary selective phase rotation
   transform for multi-qubit states. In section \ref{SPRT} we will give a short
   introduction to the quantum integral transformation and selective phase rotation
   transform. In section \ref{gen} we will recall  a theorem on the construct quantum
   gate entangler for multi-qubit state and then use this theorem to construct a quantum
    gate entangler based on selective rotation transformation. We also in detail discuss
 the construction  of such a entangler for three-qubit states.

\section{Selective phase rotation transform}\label{SPRT}
In this section we will define quantum fourier transform and
selective phase rotation transform \cite{Naka}. Let $m\in
\mathbb{N}$ and $S_{m}=\{0,1,\ldots,2^{m}\}$ be a set of integers.
Then, for any function $f:S_{m}\longrightarrow\mathbb{C}$, a
discrete integral transform $\widetilde{f}:S_{m}\longrightarrow
\mathbb{C}$ with kernel $K:S_{m}\times
S_{m}\longrightarrow\mathbb{C}$ is defined by
\begin{equation}
\widetilde{f}(y)=\sum^{2^{m}-1}_{x=0}K(x,y)f(x).
\end{equation}
In the matrix form the kernel $K$, the function $f$, and
$\widetilde{f}$ are given by $K=(K_{\mu\nu})_{0\leq \mu,\nu\leq
2^{m}-1}$, $f=(f(0),f(1),\ldots, f(2^{m}-1))^{T}$, and
$\widetilde{f}=Kf$, respectively. Moreover, if the kernel $K$ is
unitary, then the inverse transform  $\widetilde{f}\longrightarrow
f$ of a discrete integral transform exists and is given by
$f(x)=\sum^{2^{m}-1}_{x=0}K^{\dagger}(x,y)f(y)$. For example, let
$U$ be a unitary matrix acting on  the $m$-qubit space
$\mathcal{H}_{\mathcal{Q}}$ and
$\{\ket{x}=\ket{x_{m-1},x_{m-2},\ldots,x_{0}}\}_{x_{j}\in\{0,1\}}$,
be a binary basis of $\mathcal{H}$, where
$x=x_{m-1}2^{m-1}+x_{m-2}2^{m-2}+\cdots+x_{0}2^{0}$. Then,
$U\ket{x}=\sum^{2^{m}-1}_{y=0}U(x,y)\ket{y}$, where $U(x,y)=\langle
x \ket{Uy}$ is the $(x,y)$-component of the matrix $U$. Moreover,
the unitary matrix $U$ implementing a discrete integral transform as
follows
\begin{equation}
U\left(\sum^{2^{m}-1}_{x=0}f(x)\ket{x}\right)=\sum^{2^{m}-1}_{y=0}\widetilde{f}(y)\ket{y}
\end{equation}
is called the quantum integral transform.

Now, let $K_{m}(x,y)=e^{\varphi_{x}}\delta_{xy}$, for all $x,y\in
S_{m}$ and $\varphi_{x}\in \mathbb{R}$. Then the discrete integral
transform
\begin{equation}
\widetilde{f}(y)=\sum^{2^{m}-1}_{x=0}K(x,y)f(x)=\sum^{2^{m}-1}_{x=0}e^{i
\varphi_{x}}\delta_{xy}f(x) =e^{i\varphi_{x}}f(y)
\end{equation}
is called the selective phase rotation transform. Note that $K_{m}$
is a unitary transformation. For example, in matrix form $K_{1}$ and
$K_{2}$ are given by $K_{1}=\left(
                              \begin{array}{cc}
                                e^{i\varphi_{0}}& 0 \\
                                0 & e^{\varphi_{1}} \\
                              \end{array}
                            \right)$ and
\begin{equation}
K_{2}=\left(
        \begin{array}{cc|cc}
          e^{i\varphi_{0}} & 0 & 0 & 0 \\
          0 & e^{i\varphi_{1}} & 0 & 0 \\
          \hline
          0 & 0 & e^{i\varphi_{2}} & 0 \\
          0 & 0 & 0 & e^{i\varphi_{3}} \\
        \end{array}
      \right)=L_{0}L_{1},
\end{equation}
where
\begin{equation}
L_{0}=A_{0}\otimes U_{0}+A_{1}\otimes I_{2\times2}=\left(
                                          \begin{array}{c|c}
                                           U_{0} & 0_{2\times2}\\
                                           \hline
                                            0_{2\times2} & I_{2\times2} \\
                                          \end{array}
                                        \right),
\end{equation}
\begin{equation}
L_{1}=A_{0}\otimes I_{2\times2}+A_{1}\otimes U_{1}=\left(
                                          \begin{array}{c|c}
                                           I_{2\times2} & 0_{2\times2}\\
                                           \hline
                                            0_{2\times2} & U_{1} \\
                                          \end{array}
                                        \right),
\end{equation}
 $U_{0}=\left(
                              \begin{array}{cc}
                                e^{i\varphi_{0}}& 0 \\
                                0 & e^{\varphi_{1}} \\
                              \end{array}
                            \right)$,  $U_{1}=\left(
                              \begin{array}{cc}
                                e^{i\varphi_{2}}& 0 \\
                                0 & e^{\varphi_{3}} \\
                              \end{array}
                            \right)$, $A_{0}=\left(
                                        \begin{array}{cc}
                                          1 & 0\\
                                          0 & 0\\
                                        \end{array}
                                      \right)$, and $A_{1}=\left(
                                        \begin{array}{cc}
                                          0 & 0\\
                                          0 & 1\\
                                        \end{array}
                                      \right)$
                            .
Note also that $L_{0}$ and $L_{1}$ gates could be implemented with a
set of universal gates, e.g., $L_{0}$ is a controlled-$U_{0}$ gate
where the control bit is negated and $L_{1}$ is a controlled-$U_{1}$
gate.

\section{Quantum gate entangler for multi-qubit states}\label{gen}
In this section we will review the construction of  a unitary
operator that entangle a  multi-qubit state \cite{Hosh3}. Then, we
will construct quantum gate entangler for multi-qubit states based
on selective phase rotation transform. We will also in detail
discuss this operator for three-qubit states.
For a multi-qubit state $\ket{\Psi}=\sum^{1}_{x_{m-1},\ldots,
x_{0}=0}\alpha_{x_{m-1}x_{m-2}\cdots x_{0}}\ket{x}$ a geometrical
unitary transformation $\mathcal{R}_{2^{m}\times 2^{m}}$ that create
multipartite entangled state is defined by $\mathcal{R}_{2^{m}\times
2^{m}}=\mathcal{R}^{d}_{2^{m}\times
2^{m}}+\mathcal{R}^{ad}_{2^{m}\times 2^{m}}$, where
\begin{equation}
\mathcal{R}^{a}_{2^{m}\times
2^{m}}=\mathrm{diag}(\alpha_{00\cdots0},0,\ldots,0,\alpha_{11\cdots0},0,\ldots,0,\alpha_{11\cdots
1})
\end{equation}
 is a
diagonal matrix and
\begin{equation}\mathcal{R}^{ad}_{2^{m}\times 2^{m}}=\mathrm{antidiag}(0,\alpha_{11\cdots0},
\ldots,\alpha_{10\cdots0},\alpha_{01\cdots 1},
\ldots,\alpha_{0\cdots01},0)
\end{equation}
 is an anti-diagonal matrix.
 Then, we have the following theorem for a multi-qubit
state.
\begin{thm} If elements of $\mathcal{R}_{2^{m}\times 2^{m}}$ satisfy
\begin{eqnarray}\label{segreply1}
&& \alpha_{x_{m-1}x_{m-2}\cdots x_{0}}\alpha_{y_{m-1}y_{m-2}\cdots
y_{0}}\\\nonumber&&\neq \alpha_{x_{m-1}x_{m-2}\ldots
x_{j+1}y_{j}x_{j-1}\cdots x_{0}}\alpha_{y_{m-1}y_{m-2} \cdots
y_{j+1} x_{j}y_{j-1}\ldots y_{0}},
\end{eqnarray}then the state
$\mathcal{R}_{2^{m}\times
2^{m}}(\ket{\psi}\otimes\ket{\psi}\times\cdots\otimes\ket{\psi})$,
with $\ket{\psi}=\ket{0}+\ket{1}$ is entangled.
\end{thm}
The proof of this theorem is based on the Segre variety
\begin{eqnarray}\label{eq: submeasure}
&&\bigcap_{\forall j}\mathcal{V}(\alpha_{x_{m-1}x_{m-2}\cdots
x_{0}}\alpha_{y_{m-1}y_{m-2}\cdots y_{0}}\\\nonumber&&-
\alpha_{x_{m-1}x_{m-2}\ldots x_{j+1}y_{j}x_{j-1}\cdots
x_{0}}\alpha_{y_{m-1}y_{m-2} \cdots y_{j+1} x_{j}y_{j-1}\ldots
y_{0}}).
\end{eqnarray}
which is the image of the Segre embedding and is an intersection of
families of quadric hypersurfaces in
$\mathbb{P}_{\mathbb{C}}^{2^{m}-1}$ that represents the completely
decomposable tensor. Note that  we need to imposed some constraints
on the parameters $\alpha_{x_{m-1}x_{m-2}\cdots x_{0}}$ to ensure
that our $\mathcal{R}_{2^{m}\times 2^{m}}$ is unitary. Note also
that for a multi-qubit state
\begin{eqnarray}
\tau_{2^{m}\times 2^{m}}&=&\mathcal{R}_{2^{m}\times
2^{m}}\mathcal{P}_{2^{m}\times 2^{m}}\\\nonumber&=&
\mathrm{diag}(\alpha_{0\cdots0 0},\alpha_{0\cdots0
1},\ldots,\alpha_{1\cdots 11})
\end{eqnarray}
 is a $2^{m}\times 2^{m}$
phase gate and $\mathcal{P}_{2^{m}\times 2^{m}}$ is $2^{m}\times
2^{m}$ swap gate.

Now, we will construct a quantum gate entangler based on the kernel
of selective phase transform kernel for a multi-qubit state.
Let the   unitary transformation $K_{m}(x,y)$ be defined as in
section \ref{SPRT}. Moreover, let
\begin{equation}
\mathcal{K}_{2^{m}\times 2^{m}}=\frac{1}{\sqrt{2^{m}}}\left(
        \begin{array}{cccc}
          e^{i\varphi_{00\cdots0}} & 0 & \cdots & 0 \\
          0 & e^{i\varphi_{00\cdots01}} & 0 & 0 \\
          \vdots & 0 & \ddots & \vdots \\
          0 & \cdots& 0 & e^{i\varphi_{11\cdots1}} \\
        \end{array}
      \right).
\end{equation}
 Then, we also have the following theorem.
\begin{thm} If elements of $\mathcal{K}_{2^{m}\times 2^{m}}$ satisfy
\begin{eqnarray}\label{segreply1}\nonumber
&& e^{i\varphi_{x_{m-1}x_{m-2}\ldots
x_{0}}}e^{i\varphi_{y_{m-1}y_{m-2}\ldots y_{0}}}\\\nonumber&&\neq
e^{i\varphi_{x_{m-1}x_{m-2}\cdots x_{j+1}y_{j}x_{j-1}\cdots
x_{m-1}}}e^{i\varphi_{y_{m-1}y_{m-2} \cdots y_{j+1}
x_{j}y_{j-1}\cdots y_{0}}}\\ &\Longleftrightarrow&
\varphi_{x_{m-1}x_{m-2}\ldots x_{0}}+\varphi_{y_{m-1}y_{m-2}\ldots
y_{0}}\\\nonumber&&\neq \varphi_{x_{m-1}x_{m-2}\cdots
x_{j+1}y_{j}x_{j-1}\cdots x_{m-1}}+\varphi_{y_{m-1}y_{m-2} \cdots
y_{j+1} x_{j}y_{j-1}\cdots y_{0}},
\end{eqnarray}then the state
$\mathcal{K}_{2^{m}\times
2^{m}}(\ket{\psi}\otimes\ket{\psi}\times\cdots\otimes\ket{\psi})$,
with $\ket{\psi}=\ket{0}+\ket{1}$ is entangled.
\end{thm}
The proof of this theorem also follows from the decomposable tensor
in terms of the Segre variety. As an example, we will construct such
quantum gate entangler for a three-qubit state. In this case,
$K_{m}(x,y)$ is given by
\begin{equation}
\mathcal{K}_{8\times 8}=\frac{1}{\sqrt{8}}\left(
        \begin{array}{cccc}
          e^{i\varphi_{000}} & 0 & \cdots & 0 \\
          0 & e^{i\varphi_{001}} &  & 0 \\
          \vdots &  & \ddots & \vdots \\
          0 & \cdots& 0 & e^{i\varphi_{111}} \\
        \end{array}
      \right)=L^{3}_{0}L^{3}_{1}L^{3}_{2}L^{3}_{3},
\end{equation}
where $L^{3}_{0},L^{3}_{1},L^{3}_{2}$, and $ L^{3}_{3}$ are
constructed by universal controlled gates as follows
\begin{equation}
L^{3}_{0}=\left(
  \begin{array}{c|c}
    L^{2}_{0} & I_{4\times4} \\
    \hline
    I_{4\times4} & I_{4\times4} \\
  \end{array}
\right)=\left(
  \begin{array}{c|c}
  \begin{array}{c|c}
    U_{0} & 0_{2\times2} \\
    \hline
    0_{2\times2} & I_{2\times2} \\
  \end{array}
   & I_{4\times4} \\
    \hline
    I_{4\times4} & I_{4\times4} \\
  \end{array}
\right), ~ U_{0}=\left(
  \begin{array}{cc}
     e^{i\varphi_{000}}& 0 \\
    0 &  e^{i\varphi_{001}}\\
  \end{array}
\right)
\end{equation}
\begin{equation}
L^{3}_{1}=\left(
  \begin{array}{c|c}
    L^{2}_{1} & I_{4\times4} \\
    \hline
    I_{4\times4} & I_{4\times4} \\
  \end{array}
\right)=\left(
  \begin{array}{c|c}
  \begin{array}{c|c}
    I_{2\times2} & 0_{2\times2} \\
    \hline
    0_{2\times2} & U_{1} \\
  \end{array}
 & I_{4\times4} \\
    \hline
    I_{4\times4} & I_{4\times4} \\
  \end{array}
\right), ~ U_{1}=\left(
  \begin{array}{cc}
     e^{i\varphi_{010}}& 0 \\
    0 &  e^{i\varphi_{011}}\\
  \end{array}
\right)
\end{equation}
\begin{equation}
L^{3}_{2}=\left(
  \begin{array}{c|c}
    I_{4\times4} & I_{4\times4} \\
    \hline
    I_{4\times4} & L^{2}_{0} \\
  \end{array}
\right)=\left(
  \begin{array}{c|c}
  I_{4\times4}
 & I_{4\times4} \\
    \hline
    I_{4\times4} & \begin{array}{c|c}
    U_{2} & 0_{2\times2} \\
    \hline
    0_{2\times2} & I_{2\times2} \\
  \end{array} \\
  \end{array}
\right), ~ U_{2}=\left(
  \begin{array}{cc}
     e^{i\varphi_{100}}& 0 \\
    0 &  e^{i\varphi_{101}}\\
  \end{array}
\right),
\end{equation}
and
\begin{equation}
L^{3}_{3}=\left(
  \begin{array}{c|c}
    I_{4\times4} & I_{4\times4} \\
    \hline
    I_{4\times4} & L^{2}_{1} \\
  \end{array}
\right)=\left(
  \begin{array}{c|c}
  I_{4\times4}
 & I_{4\times4} \\
    \hline
    I_{4\times4} & \begin{array}{c|c}
    I_{2\times2} & 0_{2\times2} \\
    \hline
    0_{2\times2} & U_{3} \\
  \end{array} \\
  \end{array}
\right), ~ U_{3}=\left(
  \begin{array}{cc}
     e^{i\varphi_{110}}& 0 \\
    0 &  e^{i\varphi_{111}}\\
  \end{array}
\right),
\end{equation}
where $0_{2\times2}$ and $0_{4\times4}$ are 2-by-2 and 4-by-4 zero
matrices and $I_{2\times2}$ and $I_{4\times4}$ are 2-by-2 and 4-by-4
identity matrices. Following  this construction of quantum gate
entangler for three-qubit states based on universal controlled gates
we can also write
\begin{equation}
\mathcal{K}_{2^{m}\times 2^{m}}=L^{m}_{0}L^{m}_{1}\cdots
L^{m}_{2^{m-1}-1},
\end{equation}
where $L^{m}_{i}$, for $1\leq i\leq 2^{m-1}-1$ can be express in
terms of universal controlled gates, e.g.,
\begin{equation}
L^{m}_{0}=\left(
  \begin{array}{c|c}
  \begin{array}{c|c}
    U_{0} & 0_{2\times2} \\
    \hline
    0_{2\times2} & I_{2\times2} \\
  \end{array}
   & 0_{4\times2^{m-1}} \\
    \hline
    0_{2^{m-1}\times4} & I_{2^{m-1}\times2^{m-1}} \\
  \end{array}
\right), ~ U_{0}=\left(
  \begin{array}{cc}
     e^{i\varphi_{00\cdots0}}& 0 \\
    0 &  e^{i\varphi_{00\cdots01}}\\
  \end{array}
\right),
\end{equation}
where $0_{2^{m-1}\times2^{m-1}}$ and $I_{2^{m-1}\times2^{m-1}}$ are
$2^{m-1}$-by-$2^{m-1}$ zero
 and identity matrices, respectively.
Thus we have following corollary.
\begin{cor} If elements of universal controlled gates $L^{m}_{0}L^{m}_{1}\cdots
L^{m}_{2^{m-1}-1}$  satisfy
\begin{eqnarray}\label{segreply1}\nonumber
&& e^{i\varphi_{x_{m-1}x_{m-2}\ldots
x_{0}}}e^{i\varphi_{y_{m-1}y_{m-2}\ldots y_{0}}}\\\nonumber&&\neq
e^{i\varphi_{x_{m-1}x_{m-2}\cdots x_{j+1}y_{j}x_{j-1}\cdots
x_{m-1}}}e^{i\varphi_{y_{m-1}y_{m-2} \cdots y_{j+1}
x_{j}y_{j-1}\cdots y_{0}}}
\end{eqnarray} then the state
$L^{m}_{0}L^{m}_{1}\cdots
L^{m}_{2^{m-1}-1}(\ket{\psi}\otimes\ket{\psi}\times\cdots\otimes\ket{\psi})$
is entangled.
\end{cor}
In this paper we have investigated the construction of quantum gate
entangler based on selective phase rotation transform. We have shown
that this quantum gate entangler can be constructed by universal
controlled gates for multi-qubit states. we also have establish a
relation between geometrical structures of multi-qubit states and
such a quantum gate entangler. These results give new way of
constructing topological and geometrical quantum gates entangler for
multipartite quantum systems.

\end{document}